\input{psfig}
\documentstyle[12pt]{article}

\begin{document}

\textheight 23cm
\textwidth 16cm
\oddsidemargin 0in
\evensidemargin 0in
\topmargin -0.25in

\centerline{\bf\Large Charge Solitons in 1-D Arrays}
\centerline{\bf\Large of Serially Coupled Josephson Junctions}
\vskip .5cm
\centerline{\large Ziv Hermon$^{(1)}$, Eshel Ben-Jacob$^{(2)}$, 
Gerd Sch\"on$^{(1)}$}
\vskip.5cm
\noindent
$^{(1)}$Institut f\"ur Theoretische Festk\"orperphysik,
Universit\"at Karlsruhe, D-76128 Karlsruhe, Germany
\\
$^{(2)}$School of Physics and Astronomy, Raymond and Beverly
Sackler Faculty of Exact Sciences, Tel-Aviv University, 
Ramat-Aviv, 69978 Tel-Aviv, Israel

{\abstract
We study a 1-D array of Josephson coupled superconducting grains with
kinetic inductance which dominates over the Josephson inductance. 
In this limit the dynamics of excess Cooper pairs in the array is 
described in terms of charge solitons, created by polarization of the 
grains. We analyze the dynamics of these topological 
excitations, which are dual to the fluxons in a long Josephson junction,
using the continuum sine-Gordon model. We find that their 
classical relativistic motion leads to 
saturation branches in the I-V characteristic of the array. We then 
discuss the semi-classical quantization of the charge soliton, and 
show that it is consistent with the large kinetic inductance of the array. 
We study the dynamics of a quantum charge soliton in a ring-shaped array 
biased by an external flux through its center. If the dephasing length of 
the quantum charge soliton is larger than the circumference of the array, 
quantum phenomena like persistent current and coherent current oscillations
are expected. As the characteristic width of the charge soliton is of
the order of $100\,\mu m$, it is a macroscopic quantum object.
We discuss the dephasing mechanisms which can suppress the quantum 
behaviour of the charge soliton.}

\section{Introduction}

\begin{figure}
\centerline{\psfig{figure=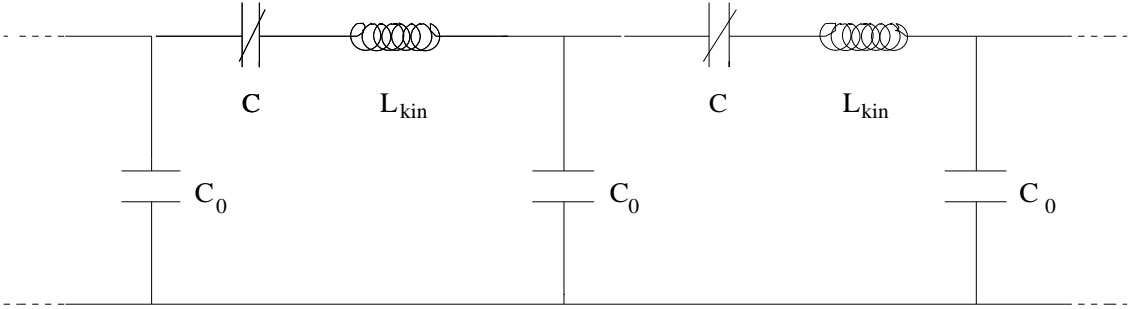,width=15.0cm}}
\caption{An equivalent electric circuit of a 1-D array of
serially coupled Josephson junctions.}
\label{FIG1}
\end{figure}

Arrays of Josephson junctions in 1-D, 2-D or 3-D have been studied 
extensively in recent years, both theoretically and experimentally 
\cite{Karlsruhe94}. When the capacitance of the junctions is small, the 
arrays are usually characterised by the Josephson energy,
$\sum_iE_J[1-\cos(\phi_i-\phi_{i+1})]$, and by the charging energy, 
${1\over2}\sum_{ij}Q_iC^{-1}_{ij}Q_j$. Here $\phi_i$ and $Q_i$ denote the 
phase and the charge on the $i'$th grain of the array, respectively, 
$C^{-1}_{ij}$ is the inverse capacitance matrix, and $E_J$ is the 
Josephson coupling energy. This description in terms of variables defined 
on the grains and not on the junctions is consistent with the fact that 
the kinetic and the geometric inductances of the grains are typically 
smaller than the Josephson inductance. As a result, the charge 
redistribution time in the grains is shorter than the tunneling time. In 
this paper we study the opposite limit, which is also experimentally 
accessible, namely a 1-D array where the kinetic inductance of the grains
\begin{equation}
\label{KIN_IND}
L_{kin}={m^*_el_x\over {e^*}^2n_sS}\ ,
\end{equation}
dominates over the Josephson inductance
\begin{equation}
\label{L_J}
L_J={1\over (2\pi)^2}{\Phi_0^2\over E_J}\ .
\end{equation}
Here $m^*_e$ and $e^*$ are the Cooper pair mass and charge, $n_s$ the 
Cooper pairs density, $l_x$ the length of a grain and $S$ the cross 
section of a grain. The large kinetic inductance means that in this case 
the charge redistribution time in the grains is longer than the tunneling 
time, thus the dynamic variables should be defined on the junctions of 
the array and not on the grains. This array can be represented by the
electric circuit shown in Fig. \ref{FIG1}. $C_0$ denotes the
self-capacitance of the superconducting grains, while the combined effect
of the Josephson and charging energies of the junctions results in a 
non-linear capacitance, $C$, as we explain in the next section.
We show that in this kind of array the concept of 'charge soliton'
\cite{BJMA}-\cite{HAVILAND} arises, 
i.e. an excess Cooper pair in the array gives rise to a compact 
topological solitonic excitation. This appears to be in contrast to the 
usual model which does not incorporate the inductive effects. That model 
suggests that an excess Cooper pair delocalises as a consequence of the 
Josephson tunneling. We show, however, that a sufficiently large kinetic 
inductance decouples the individual junctions quantum mechanically. We 
study the dynamics of the charge soliton both classically and quantum 
mechanically.

The paper is organised as follows: In section (2) we develop a continuum 
approximation of a serially coupled array of Josephson junctions with a
dominant kinetic inductance. In section (3) we show that this array has
compact solitonic excitations ('charge solitons'), and discuss some of 
their classical
properties and dynamics. In this section we discuss the small amplitude 
oscillations of the array ('plasmons') as well. In section (4) we study 
the classical dynamics of the charge soliton further, using collective 
coordinates. 
The quantization of the charge soliton is done in section (5). We discuss
the meaning of the semi-classical quantization of the soliton, and study 
its quantum dynamics in a ring-shaped array. We demonstrate that quantum
charge solitons can, in principle, exhibit quantum phenomena without 
classical analogues, like persistent motion in response to an external 
flux and coherent current oscillations. We then discuss possible dephasing 
mechanisms of charge solitons, and address the effects caused by the 
discreteness of the array. We summarize our results in the concluding 
section (6).

\section{Kinetic Inductance Dominated 1-D Array of Serially Coupled 
Josephson Junctions}

\subsection{The Lagrangian}

We consider a chain of $N$ identical superconducting grains (thus
forming $N-1$ Josephson junctions). The junctions are characterised by the
Josephson coupling energy and by the charging energy scale
\begin{equation}
\label{E_C}
E_C\equiv {(2e)^2\over 2C}\ .
\end{equation}
We assume that $C\approx 10^{-15}\,$Farad, and that $E_J$ is of the same 
order as $E_C$.
The grains are capacitively coupled to a conducting substrate 
with a capacitance $C_0\ll C$, which we assume to be 
$C_0\approx 10^{-17}\,$Farad. The energy scale of this coupling energy, 
\begin{equation}
\label{E_C_0}
E_{C_0}\equiv{(2e)^2\over 2C_0}\ ,
\end{equation}
is thus much larger than the junction charging energy
\begin{equation}
\label{E_C_0_E_C}
E_{C_0}\gg E_C\ .
\end{equation}
The grains are characterised by the inductive energy scale associated with 
$L_{kin}$ 
\begin{equation}
\label{E_L}
E_L\equiv{\Phi_0^2\over2L_{kin}}\ ,
\end{equation}
where $\Phi_0\equiv h/2e$. 
As we have said in the introduction, we assume 
that the kinetic inductance of the grain dominates over the Josephson 
inductance. In fact, due to the numerical coefficient $(2\pi)^2/2$ 
difference in the relations (\ref{L_J}) and (\ref{E_L}), 
$L_{kin}$ should be larger than $2\pi^2L_J$ for the
inductive effects to be important. For a typical $E_J$ of 
the order of $100\,\mu eV$ it means that $L_{kin}$ dominates if it is 
$10^{-7}\,$Henry or larger. This situation can be achieved, for instance, 
when $l_x\approx 10\,\mu m$ and $S\approx 10^3\, nm^2$. Nevertheless we 
assume that the width of the grains is of the order of the London 
penetration depth to avoid tunneling of flux quanta through the grains.
The width of the junctions, $d$, is much smaller than $l_x$ (typically 
$d\approx 2\, nm$), and the 
distance between adjacent grains (the unit cell) 
is denoted by $a$ ($a\equiv l_x+d$). $L\equiv Na$ is the total length 
of the chain. We assume that the chain is very long ($N\gg 1$). 

Using the values given above, we find that the zero frequency impedance 
of a unit cell, 
\begin{equation}
\label{IMP}
Z_{LC}=\sqrt{L_{kin}/C_0}\ ,
\end{equation}
is of the order of $100\,K\Omega$, i.e. it is much larger than the quantum
resistance, $R_Q\equiv h/(2e)^2$:
\begin{equation}
\label{IMP_CON}
Z_{LC}\gg R_Q\ .
\end{equation}
Note that this impedance inequality can be expressed alternatively as an
inequality of the coupling energy and the inductive energy scales 
\begin{equation}
\label{E_L_E_C_0}
E_{C_0}\gg E_L\ .
\end{equation}
A similar condition to (\ref{IMP_CON}) has been studied before in the 
context of single electron tunneling in a normal junction \cite{Dec_Env}, 
and it has been shown that it leads to a quantum mechanical decoupling of 
the junction from its environment. Using the same reasoning here, we are 
led to the conclusion that the condition (\ref{IMP_CON}) means that each 
junction is quantum mechanically decoupled from its environment, i.e. from 
the other junctions of the array. We can thus solve the Schr\"odinger 
equation for each junction separately, and obtain a local potential energy 
of the array. This situation has been named the 'local rule' in the context
of single electron tunneling \cite{Loc_Glob}. 

\begin{figure}
\centerline{\psfig{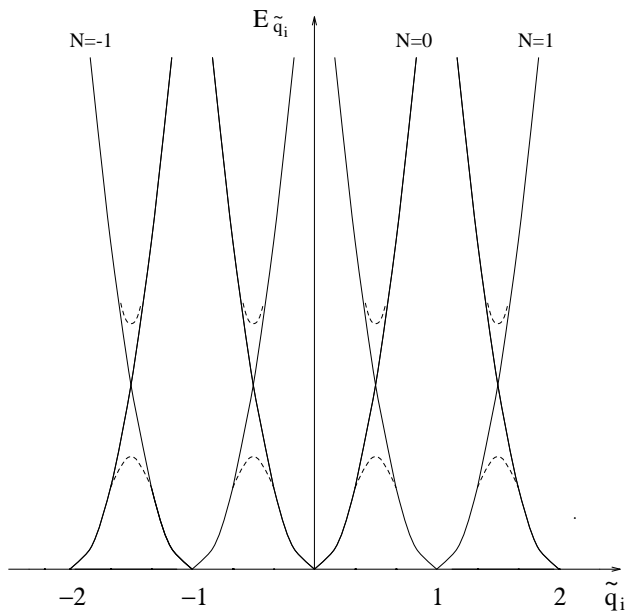}}
\caption{Energy levels of a Josephson junction as a function of 
$\tilde q_i$ for the case $E_J\approx E_C$.}
\label{FIG2}
\end{figure}

The eigenstates of the junction 
$i$ depend on $\tilde q_i$, the dimensionless charge (in units of $2e$) 
induced on this junction. As a function of $\tilde q_i$, the energy levels 
are made of a set of charging energy parabolas, with gaps at the 
intersection regions due to the Josephson energy 
\cite{Widom_Bloch}-\cite{Gerd_Andrei} (see Fig. \ref{FIG2}). 
The energy levels are, thus, periodic functions of $\tilde q_i$ with a 
period $1$. Under appropriate conditions (not too small gaps, adiabatic 
changes) Zener transitions between the levels can be avoided 
\cite{BJGMS_SQ}, \cite{BJGMS}.
We also ignore, for the time being, quasi-particle tunneling, which is a 
dissipative process. We discuss this issue in section (5). We thus may
consider only the first level, which we denote by $E_{\tilde q_i}$. 
This level represents coherent superposition of charge states in the bulk
superconductors, which differ by one Cooper pair. 
$E_{\tilde q_i}$ is formally given as an eigenvalue of Mathieu's equation. 
As it does not have a simple analytical form when $E_C$ is of the same 
order of $E_J$, and our results do not depend qualitatively on the exact 
form of $E_{\tilde q_i}$, we adopt the following form
\begin{equation}
\label{7_ARRAY_PER_NLIN_CH_ENE}
E_{\tilde q_i}={2\over(2\pi)^2}E_C[1-\cos(2\pi\tilde q_i)]\ .
\end{equation}
The form (\ref{7_ARRAY_PER_NLIN_CH_ENE}) preserves the correct parabolic 
dependence for small $q_i$, and reduces the amplitude of the energy level 
from its maximal height (in the limiting case where $E_J=0$) by a factor 
of $\pi^2/4$. We emphasize that the important feature of 
$E_{\tilde q_i}$ is its periodicity, which allows us to represent the 
Josephson junction as a non-linear capacitor (see Fig. \ref{FIG1}). In 
the next 
section we show that the periodicity gives rise to the soliton description.

Due to the tunneling of Cooper pairs the variable $\tilde q_i$ is compact, 
i.e. $\tilde q_i+1=\tilde q_i$. It is convenient to introduce an 
extended variable $q_i$, which is the dimensionless charge (in units of 
$2e$) brought to the i'th junction. $q_i$ is related to $\tilde q_i$ 
through 
\begin{equation}
\label{7_DIS_q_DEF}
q_i\equiv \tilde q_i+\sum_{i'=i+1}^NQ_{i'} \ ,
\end{equation}
where $Q_i$ is the net charge on the $i$'th {\it grain}. $Q_i$ has, of 
course, only discrete values, while $q_i$ and $\tilde{q_i}$ are continuous.
This change of variables corresponds to changing from a 'reduced
zone' scheme to an 'extended zone' scheme in the junction's energy bands 
(Fig. \ref{FIG2}). 
This variable was used in the study of 1-D array of serially coupled normal
junction as well \cite{BJMA}, \cite{TOP_EX}. In the next section we show 
the importance of $q_i$ for the solitonic description.
The form of the energy of the junction (the potential energy) does not 
change when expressed as a function of $q_i$
\begin{equation}
\label{7_ARRAY_NLIN_CH_ENE}
E_{pot}=E_{q_i}={2\over(2\pi)^2}E_C[1-\cos(2\pi q_i)]\ .
\end{equation}
The voltage across the junction, $V_{q_i}$, is given by the derivative 
of the energy levels with respect to the charge
\begin{equation}
\label{V_q}
V_{q_i}={1\over 2e}{\partial E_{q_i}\over\partial q_i}\ .
\end{equation}
Using (\ref{7_ARRAY_NLIN_CH_ENE}) we express the voltage as
\begin{equation}
\label{V_Q}
V_{q_i}={1\over 2\pi}V_C\sin 2\pi q_i\ ,
\end{equation}
where $V_C\equiv{2e\over C}$.

Since $q_i$ is defined on the junction it is already contains an averaging
over the fast tunneling process. A time dependent $q_i$ is therefore 
related to the slow process of charge redistribution in the grains by 
means of a supercurrent. This gives rise to an inductive energy in the 
grains, which serves as the kinetic energy of the array:
\begin{equation}
\label{7_ARRAY_IND_ENE}
E_{kin}={1\over 2}(2e)^2L_{kin}\dot q_i^2\ .
\end{equation}
In the parameters range we consider, the kinetic energy scale is smaller 
than the potential energy one
\begin{equation}
\label{E_C_E_L}
E_C>E_L\ .
\end{equation}
The three inequalities: (\ref{E_C_0_E_C}), (\ref{E_L_E_C_0}) and 
(\ref{E_C_E_L}) can be combined into a single condition for the 
energy scales of the system 
\begin{equation}
\label{COM_SCALE_REL}
E_{C_0}>E_C>E_L\ .
\end{equation}

The relation between the dynamic variable $q_i$ and the voltage $V_i$ 
between the $i$'th grain and the substrate can be found by consecutive 
applications of Gauss' law
\begin{equation}
\label{7_DIS_q_V}
q_i=q_1-{1\over 2e}C_0\sum_{i'=1}^iV_{i'} \ ,
\end{equation}
where $q_1$ is the charge that was brought to the first junction of the 
array. From
now on we assume that the continuum limit can be taken. (We will show the 
necessary condition for this soon.) Discreteness effects are discussed in 
section (5). In the continuum limit Eqs. (\ref{7_DIS_q_DEF}) and 
(\ref{7_DIS_q_V}) have the form 
\begin{equation}
\label{7_q_DEF}
q(x)\equiv \tilde q(x)+\int_x^{L} Q(\xi)\,d\xi/a \ ,
\end{equation}
\begin{equation}
\label{7_q_V}
q(x)=q(0)-{1\over 2e}C_0\int_{0}^xV(\xi)\,d\xi/a \ .
\end{equation}
The array is thus described by the charge field $q(x)$.
The relation (\ref{7_q_V}) between $q(x)$ and $V(x)$ can be expressed in a
local form
\begin{equation}
\label{7_LOC_q_V}
V(x)=-a{2e\over C_0}q_x(x)\ .
\end{equation}
We see that the $q_x(x)$ is the dimensionless charge between the grains
and the substrate.
The charging energy which couples the unit cells of the array can be
expressed, therefore, as
\begin{equation}
\label{7_COUPLING_ENE}
E_{coupling}=a^2{(2e)^2\over 2C_0}q_x^2 \ .
\end{equation}
As we have mentioned above, its energy scale is $E_{C_0}$ (\ref{E_C_0}).
When $C_0\ll C$ we have $E_{C_0}\gg E_C$. In this case even
small amounts of charge induce high voltages on the capacitors between the
grains and the substrate, and these voltages strongly couple the Josephson
junctions. In the opposite case when $C_0$ is large, there is almost no
voltage on the capacitors and the junctions are practically decoupled.
A small $C_0$ is thus needed for the picture of serially coupled Josephson
junctions.

{}From the above discussion we conclude that the array we consider is
characterised by the three energies: the potential energy 
(\ref{7_ARRAY_NLIN_CH_ENE}), the kinetic (or inductive) energy 
(\ref{7_ARRAY_IND_ENE}) and the coupling (or charging) energy
(\ref{7_COUPLING_ENE}). When these three energies are combined, we get the 
following sine-Gordon Lagrangian
\begin{equation}
\label{7_BARE_LAGRANGIAN}
{\cal L}={1\over 2a}(2e)^2L_{kin}\dot q^2
-a{(2e)^2\over 2C_0}q_x^2-
{1\over a}{2\over (2\pi)^2}{(2e)^2\over 2C}[1-\cos(2\pi q)]\ .
\end{equation}
This is a novel description of a 1-D Josephson junctions array, which is 
valid when the condition (\ref{COM_SCALE_REL}) holds. The three effects of 
the large kinetic inductance are reflected in the Lagrangian 
(\ref{7_BARE_LAGRANGIAN}):
1. An additional inductive energy, which is an inertial term. 2. A 
representation of each junction by a
periodic charging energy, as a result of the quantum mechanical decoupling 
of the junctions. 3. A description of the array by degrees of freedom 
which are defined on the junctions and not on the grains. The Lagrangian 
(\ref{7_BARE_LAGRANGIAN}) is electromagnetically dual to the Lagrangian
representing a long Josephson junction. The latter case can be understood 
as the continuum version of an array of parallely coupled Josephson 
junctions. Interchanging parallel coupling with series coupling and 
inductors with capacitors one gets the Lagrangian of the serially coupled 
Josephson junctions. Note, especially that the periodic inductive energy 
in the long Josephson junction (i.e. the Josephson energy) is replaced 
here by the periodic charging energy.

\subsection{The Equation of Motion and the Hamiltonian}

Following the standard sine-Gordon treatment \cite{COLEMAN}, \cite{Raj}, 
we redefine the charge field: $q(x)\rightarrow q'(x)\equiv q(x)/2\pi$, and 
express the Lagrangian (\ref{7_BARE_LAGRANGIAN}) as
\begin{equation}
\label{7_LAG_BETA}
{\cal L}={\hbar v_C\over 2\pi\beta^2}\left[{1\over 2v_C^2}\dot q^2
-{1\over 2}q_x{}^2-{1\over\Lambda_C^2}(1-\cos q)\right] \ .
\end{equation}
The three bulk parameters: $C$, $L_{kin}$ and $C_0$ are replaced in 
(\ref{7_LAG_BETA}) by $\Lambda_C$, $v_C$ and $\beta^2$. Here
\begin{equation}
\label{7_LAMBDA_C}
\Lambda_C\equiv a\sqrt{C\over C_0} \ ,
\end{equation}
is the characteristic length of the system. 
The condition needed for the validity of the continuum limit is therefore
\begin{equation}
\label{7_CON_LIM_LAM}
\Lambda_C\gg a\ ,
\end{equation}
or
\begin{equation}
\label{7_CON_LIM_CAP}
C\gg C_0\ ,
\end{equation}
which is consistent with the limit (\ref{E_C_0_E_C}). This is another 
manifestation of what we have discussed above: a small $C_0$ 
implies a large coupling, hence a large $\Lambda_C$. Using the values given
above we get $\Lambda_C\approx 100\,\mu m$. 
The second parameter in the Lagrangian (\ref{7_LAG_BETA}),
\begin{equation}
\label{7_WAVE_VELOCITY}
v_C\equiv {a\over\sqrt{L_{kin}C_0}}\ ,
\end{equation}
is the wave velocity of the system. It is of the order of 
$10^{-1}-10^{-2}\,c$, where $c$ is the vacuum light velocity. It is 
related to $\Lambda_C$ via the characteristic frequency 
\begin{equation}
\label{7_OMEGA_REL}
\omega_C={v_C\over\Lambda_C}=\sqrt{1\over L_{kin}C}\ ,
\end{equation}
which is of the order of $10^{11}\,$sec$^{-1}$.
The third parameter in the Lagrangian (\ref{7_LAG_BETA}),
\begin{equation}
\label{7_BETA_SCALE}
\beta^2\equiv {2\pi\hbar v_C C_0\over (2e)^2 a}\ ,
\end{equation}
sets the energy scale of the system. It does not affect the classical 
equation of motion, but its value is important in determining whether the 
system behaves classical or quantum mechanically. We return to this point 
in section (5), where we discuss the quantum dynamics of the system.

The equation of motion derived from the Lagrangian (\ref{7_LAG_BETA}) is
\begin{equation}
\label{7_ARRAY_EQ_MOT}
{1\over v_C^2}\ddot q-q_{xx}+{1\over\Lambda_C^2}\sin q=0 \ .
\end{equation}
It is a voltage equation for the junction, as can be shown more clearly by
multiplying it by $2ev_C^2L_{kin}/2\pi$ and using Eq. (\ref{7_LOC_q_V}) to
obtain
\begin{equation}
\label{7_VOL_EQ_MOT}
{1\over 2\pi}2eL_{kin}\ddot q-{1\over 2\pi}a^2{2e\over C_0}q_{xx}+
{1\over 2\pi}V_C\sin q=0 \ .
\end{equation}
The first term is an inductive voltage induced along the grains when the
current is time-dependent. From Eq. (\ref{7_LOC_q_V}) we see that the 
second term is the continuum form of $V_{i+1}-V_i$, i.e. it is the 
difference of the voltages between two adjacent cells and the substrate. 
The third term is the voltage across the junctions, resulting from the 
superposition of charge states (Eq. (\ref{V_Q})). The voltage equation 
(\ref{7_VOL_EQ_MOT}) is thus a Kirchoff's law for a closed loop of the 
equivalent electrical circuit of the array (Fig. \ref{FIG1}). 
The conjugate momentum of the field $q$
\begin{equation}
\label{7_FIELD_MOM}
\pi_q\equiv{\partial{\cal L}\over \partial\dot q}=
{1\over a}\left({2e\over 2\pi}\right)^2L_{kin}\dot q=
\equiv\hbar\tilde n_{\Phi_0}\ ,
\end{equation}
is the number of flux quanta per unit length that has tunneled through
the junctions of the array. 
Using $\tilde n_{\Phi_0}$ we get the Hamiltonian of of the system
\begin{equation}
\label{7_HAM_BETA}
H=\hbar v_C\int\left\{2\pi\beta^2{1\over2}\tilde n_{\Phi_0}^2+
{1\over 2\pi\beta^2}\left[{1\over2}q_x^2+{1\over\Lambda_C^2}(1-\cos q)
\right]\right\}\,dx\ .
\end{equation}

When the array is coupled to an external voltage, $V_{ext}$, the equation
of motion (\ref{7_ARRAY_EQ_MOT}) changes to
\begin{equation}
\label{7_SER_BIAS_EQ_MOT}
{1\over v_C^2}\ddot q-q_{xx}+{1\over\Lambda_C^2}\sin q=2\pi{1\over a^2}
{C_0\over 2e}V_{cell} \ ,
\end{equation}
where 
\begin{equation}
\label{V_CELL}
V_{cell}\equiv{a\over L}V_{ext}
\end{equation}
is the part of the external voltage that is distributed on one unit cell.
Equation (\ref{7_SER_BIAS_EQ_MOT}) represents, alternatively, the case
where the array has a shape of a ring and an external flux is applied
through its center. In this case $V_{ext}\equiv -\dot\Phi_{ext}$ is the 
electromotiv force acting on the array. The flux source has, of course,
the advantage that the effects of the leads are eliminated. In any case, 
equation (\ref{7_SER_BIAS_EQ_MOT}) can be derived from the following 
Hamiltonian
\begin{equation}
\label{7_BIAS_HAMILTONIAN}
H=\hbar v_C\int\left\{2\pi\beta^2{1\over2}(\tilde n_{\Phi_0}-
\tilde n_{\Phi_{ext}})^2+
{1\over 2\pi\beta^2}\left[{1\over2}q_x^2+{1\over\Lambda_C^2}(1-\cos q)
\right]\right\}\,dx\ .
\end{equation}
In the case of a voltage source $\tilde n_{\Phi_{ext}}$ is defined as the 
integral of the external voltage per unit length and unit flux
\begin{equation}
\label{7_ARRAY_PHI_EXT}
\tilde n_{\Phi_{ext}}\equiv-{1\over L\Phi_0}\int V_{ext}\,dt\ ,
\end{equation}
while in the case of a flux source it is simply the dimensionless flux 
density. The external source thus appears in the Hamiltonian as a 
time-dependent gauge potential, in analogy to the external current in the 
long Josephson junction Hamiltonian \cite{PER_VOL}. The gauge nature of the
external voltage gives rise to the following shift of the conjugate 
momentum
\begin{equation}
\label{7_BIAS_ARRAY_MOM}
\hbar\tilde n_{\Phi_0}={1\over a}\left({2e\over 2\pi}\right)^2L_{kin}\dot q
+\hbar\tilde n_{\Phi_{ext}}\ .
\end{equation}

Dissipation processes in the system produce additional
$q$ - dependent voltage drops. 
Ohmic dissipation can be represented phenomenologically by adding to
each unit cell a resistor connected to the other elements in this cell in 
series. In this case the voltage equation (\ref{7_SER_BIAS_EQ_MOT}) becomes
\begin{equation}
\label{7_SER_BIAS_RES_EQ_MOT}
{1\over v_C^2}\ddot q+{1\over a^2}C_0R\dot q-q_{xx}
+{1\over\Lambda_C^2}\sin q=2\pi{1\over a^2}{C_0\over 2e}V_{cell} \ .
\end{equation}
This representation, which was named the `serially resistive junction' 
(SRJ) in \cite{BJMA}, is the analogue of the RSJ model \cite{McCumber},
\cite{Stewart}.

\section{Charge Solitons and Plasmons}

\subsection{A Static Charge Soliton}

Since the 1-D array of serially coupled Josephson junctions
can be described by a sine-Gordon Lagrangian (\ref{7_LAG_BETA}), we expect
that it has solitonic excitations, i.e. compact, stable topological 
configurations. Using the definition of $q$ as an
extended variable (\ref{7_q_DEF}), we observe that $q(x)$ and $q(x+2\pi)$ 
can be distinguished if there is an excess or a deficiency of Cooper pairs
in intermediate grains. The one soliton excitation represents the
charging of the junctions (or the polarization of the grains) due to an
excess Cooper pair in the array, and is called a `charge soliton'. This
term was coined in \cite{BJMA} in the context of a 1-D array of normal 
tunnel junctions. Recently, charge solitons in a 1-D array of SQUID's were 
studied experimentally \cite{HAVILAND}, and a zero current state below a 
threshold voltage was found. This voltage was interpreted as an injection 
voltage for a charge soliton. 

\begin{figure}
\centerline{\psfig{figure=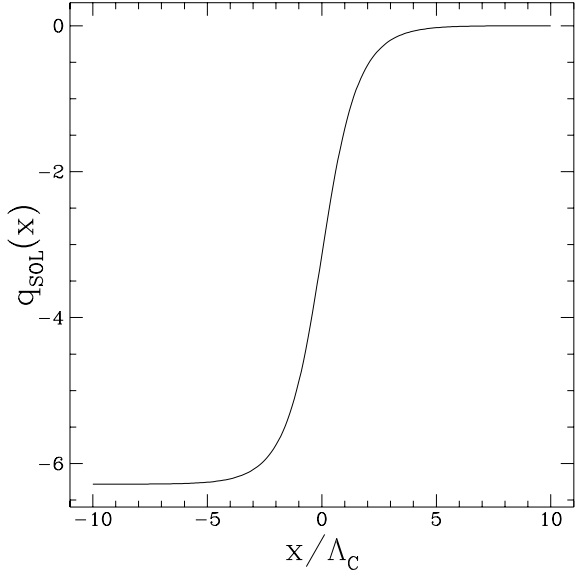,height=8.0cm}}
\caption{The charge soliton configuration representing an excess Cooper
pair in the array. The center of the soliton is taken to be $X_0=0$.}
\label{FIG3}
\end{figure}

The charge soliton solution of Eq. (\ref{7_ARRAY_EQ_MOT}) with the 
appropriate boundary conditions is (see Fig. \ref{FIG3})
\begin{equation}
\label{7_CHR_SOL}
q_{sol}(x)=
4\tan^{-1}\left[\exp\left({x-X_0\over\Lambda_C}\right)\right]-2\pi\ .
\end{equation}
Its center is at $X_0$, which we take in this section to be zero. The
excess charge of the Cooper pair is the topological charge of this soliton
\begin{equation}
\label{7_CHR_SOL_TOP_CHR}
Q=2e\int\partial_x q_{sol}\,dx=-2e\ .
\end{equation}
We would like to emphasize once more that under the condition 
(\ref{COM_SCALE_REL}) 
we consider here, the existence of a topological 
solitonic excitation and its stability do not depend on the exact form 
of the potential energy of the junctions, but only on its having 
degenerate minima. Thus our qualitative results are valid for other forms 
of the potential as well.

As was mentioned above, charge solitons in 1-D arrays of normal tunnel 
junctions have been studied previously \cite{BJMA}-\cite{DEL_SOL}. 
In this context a question was raised whether a charge soliton can be
regarded as a coherent dynamic object whose equation of motion contains 
an inertial term, as was proposed in Refs. \cite{BJMA}, \cite{ABJM} and
\cite{TOP_EX}, or that it merely represents a static charge distribution 
profile, as was argued in Refs. \cite{LIK_SOL_1} and \cite{LIK_SOL_2}. 
Here we have shown that this question should not rise in 
the Josephson junction array context. The coherence of the charge soliton 
ensues from the coherent superconducting ground state, and the inertia term
comes from the kinetic inductance of the grains. Moreover, we have 
shown that the impedance condition (\ref{IMP_CON}) should be met in order
that the concept of a charge soliton will be different from that of a point
charge (be it a Cooper pair or an electron). 

\begin{figure}
\centerline{\psfig{figure=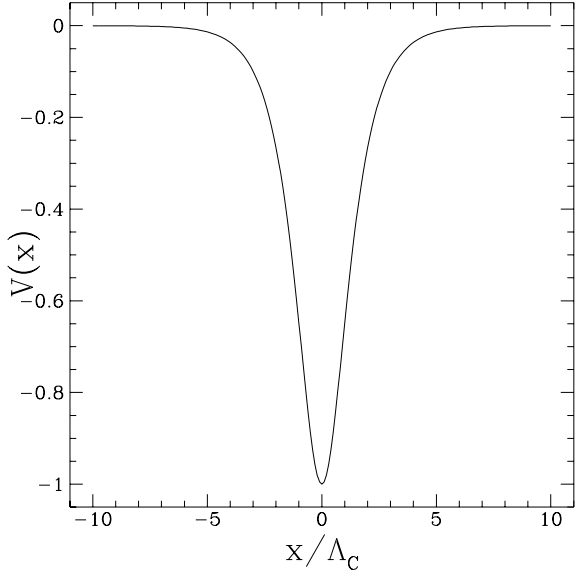,height=8.0cm}}
\caption{The profile of the voltage between the array and the
substrate induced by the charge soliton. $V$ is measured in mV.}
\label{FIG4}
\end{figure}

\begin{figure}
\centerline{\psfig{figure=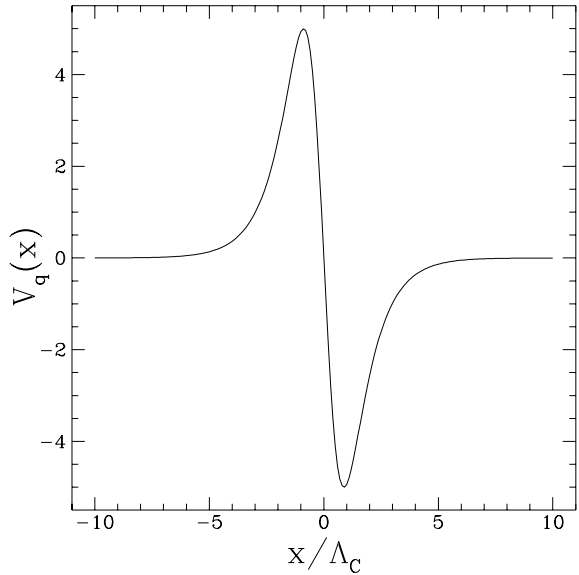,height=8.0cm}}
\caption{The distribution of voltages on the junctions of the array
corresponding to a charge soliton configuration. $V_q$ is measured in mV.}
\label{FIG5}
\end{figure}

{}From Eq.(\ref{7_CHR_SOL}) we see that the characteristic length scale of 
the array, $\Lambda_C$, is the characteristic width of the soliton as well.
In order to interpret the charge soliton as a particle its width should be 
much smaller than the total length of the array, i.e.
\begin{equation}
\label{PART_INT}
L\gg\Lambda_C\ .
\end{equation}
This assumption is met when $L\ge 10^3\, \mu m$. Here we assume that
$L\approx 10^3\, \mu m$. 
The number of grains the soliton is spread over is
\begin{equation}
\label{7_N_C}
N_C\equiv\Lambda_C/a=\sqrt{C/C_0}\ .
\end{equation}
$N_C$ is larger than one due to the continuum limit condition 
(\ref{7_CON_LIM_LAM}). For the parameters given above $N_C=10$. when the 
condition (\ref{7_CON_LIM_LAM}) fails, one should take into account 
corrections to continuum sine-Gordon model. We address this point in
section (5). The finite width of the charge
soliton is clearly seen from its density, which according to Eq. 
(\ref{7_LOC_q_V}), is proportional to the profile of the voltage 
between the array and the substrate (see Fig. \ref{FIG4})
\begin{equation}
\label{7_CHR_SOL_DEN}
V(x)=-{a\over 2\pi}{2e\over C_0}\partial_x q_{sol}(x)=-{2\over 2\pi}
{1\over N_C}{2e\over C_0}\mbox{sech}\left({x\over\Lambda_C}\right)\ .
\end{equation}
$\Lambda_C$ sets the scale for the static distribution of voltages on the 
junctions of the array as well. Using Eqs. (\ref{V_Q}) and 
(\ref{7_ARRAY_EQ_MOT}) we find that this distribution is proportional to 
the second derivative of the soliton configuration (see Fig. \ref{FIG5})
\begin{equation}
\label{CHR_SOL_SEC_DER}
V_q(x)={1\over 2\pi}V_C\Lambda_C^2\,\partial_{xx}q_{sol}(x)=
-{2\over 2\pi}V_C\,\,\mbox{sech}\left({x\over\Lambda_C}\right)
\mbox{tanh}\left({x\over\Lambda_C}\right)\ .
\end{equation}

The energy needed to create a charge soliton is the value of the 
Hamiltonian calculated for a static solution (Eq. (\ref{7_CHR_SOL}))
\begin{equation}
\label{7_SOL_ENE}
E_0={8\over\Lambda_C}{\hbar v_C\over 2\pi\beta^2}={8\over(2\pi)^2}
{(2e)^2\over\sqrt{CC_0}}={16\over(2\pi)^2}E_CN_C\ .
\end{equation}
This rest energy depends on $C$ and $C_0$ but not on $L_{kin}$, since it 
is determined by the potential and coupling energies. It can be written as 
the potential energy density ($\epsilon_C\equiv E_C/S$), times the 
effective area of the soliton ($S_{eff}\equiv SN_C$) 
\begin{equation}
\label{7_AL_SOL_ENE}
E_0={16\over(2\pi)^2}\epsilon_CS_{eff}\ .
\end{equation}
Dividing Eq. (\ref{7_SOL_ENE}) by $v_C^2$ we get the soliton rest mass
\begin{equation}
\label{7_SOL_MASS}
M_0\equiv E_0/v_C^2=
{8\over(2\pi)^2}(2e)^2{L_{kin}\over a}{1\over\Lambda_C}\ .
\end{equation}
In analogy to the rest mass of a fluxon in a long Josephson junction 
\cite{PER_VOL}, the charge soliton's rest mass is proportional to the 
inductance per unit length and inversely proportional to the characteristic
length, $\Lambda_C$. Using the typical parameters we find that the charge 
soliton mass is of the order of $10^{-36}\,$Kg, i.e. six orders of 
magnitude less than the electron rest mass. This result indicates that the 
charge soliton should not be understood as a Cooper pair dressed with a 
polarization cloud, but as the polarization cloud itself. We return to 
this point when we discuss the dynamics of the charge soliton in the next 
section.

\subsection{A Dynamic Soliton}

In order to describe a charge soliton moving with a velocity $v$, we make 
use of the Lorentz 
invariance of the Lagrangian (\ref{7_LAG_BETA}) to perform a Lorentz 
transformation of the static configuration (\ref{7_CHR_SOL}) and obtain
\begin{equation}
\label{7_DYN_SOL}
q_{sol}(x,t)=q_{sol}[\gamma(x-vt)]=4\tan^{-1}\left\{\exp\left[\gamma
\left({x-x_0-vt\over\Lambda_C}\right)\right]\right\}\ ,
\end{equation}
where $\gamma\equiv 1/\sqrt{1-(v^2/v_C^2)}$.
We thus expect that a relativistic charge soliton suffers a Lorentz
contraction. Since the light velocity in the array, $v_C$, is smaller than
the vacuum light velocity, relativistic effects of the charge soliton can 
be observed more easily than relativistic effects of electrons or Cooper
pairs.

A moving charge soliton induces, of course, a current along the array. 
The spatial distribution of the current is given by
\begin{equation}
\label{7_CURRENT}
I(x)={2e\over 2\pi}\dot q_{sol}(x)\ .
\end{equation}
This is a current pulse with a width $\Lambda_C$, concentrated around 
the moving center of the charge soliton. It has the same profile as the
voltage between the array and the substrate (see Fig. \ref{FIG4}).
The average current produced by the moving soliton is
\begin{equation}
\label{AVE_CURRENT}
\bar I={1\over L}\int I(x)\,dx=-{1\over L}2ev\ .
\end{equation}
For a soliton moving with a velocity $10^6$\, m/sec, it is of the order of
$0.1\,$nA.

\subsection{Plasmons}

Besides topological solitons, the sine-Gordon Lagrangian (\ref{7_LAG_BETA})
admits small amplitude excitations. Their dynamics is governed
by the linearized equation of motion 
\begin{equation}
\label{7_PL_EQ_MOT}
{1\over v_C^2}\ddot q-q_{xx}+{1\over\Lambda_C^2}q=0 \ .
\end{equation}
As this equation describes electromagnetic field oscillations with a 
confining potential, its solutions are longitudinal plasma oscillations 
(`plasmons') propagating along the array. The propagation of the plasmons 
does not involve any tunneling process. The plasmons have the dispersion 
relation
\begin{equation}
\label{7_DIS_REL}
\omega^2=\omega_C^2+v_C^2k^2 \ ,
\end{equation}
i.e. there is an energy gap $\hbar\omega_C$ in their spectrum with the 
corresponding temperature $T_g\approx 1\,K$. The plasmons have, therefore,
a mass 
\begin{equation}
\label{7_PLAS_MASS}
M_P={\hbar\omega_C\over v_C^2}={\hbar\over\Lambda_C v_c}\ .
\end{equation}
which is of the order of $10^{-37}\,$Kg. The ratio between the mass of a
plasmon to the mass of the soliton (\ref{7_SOL_MASS}) is $2\pi\beta^2/8$,
i.e. it is of the order of $\beta^2$. 

Plasmons can also be excited when there is a soliton in the array. In that 
case they can be considered as vibrations of the soliton. Their analytical 
form can 
be found by expanding $q$ around the soliton solution (\ref{7_CHR_SOL})
\begin{equation}
\label{2_EXP_AR_SOLITON}
q(x,t)=q_{sol}(x)+\psi_k(x)\exp(i\omega_kt)\ .
\end{equation}
Substituting (\ref{2_EXP_AR_SOLITON}) in the equation of motion 
(\ref{7_ARRAY_EQ_MOT}), and
linearizing with respect to $\psi$ one gets
\begin{equation}
\label{2_PLAS_AR_SOLITON}
\psi_k(x)\sim\left(\tanh{x\over\Lambda_C}-ik\Lambda_C\right)\exp(ikx)\ .
\end{equation}
The dispersion relation is the same as above (Eq. (\ref{7_DIS_REL})), but 
there exists now an additional zero mode (whose $\omega=0$). It reflects 
the translational invariance of the system, i.e. the homogeneity of the 
array (at distances larger than $a$). As a consequence, the zero mode is 
proportional to the spatial derivative of the soliton configuration.

\section{Collective Coordinates for the Charge Soliton}

\subsection{Definition and Equations of motion}

The topological stability of the charge soliton 
and its finite width allow for its interpretation as a particle. Thus we
would like to describe the charge soliton by a pair of conjugate
coordinates which correspond to its center of mass, $X$, and to its
momentum, $P$. This can be done by using the `collective coordinates' 
method. This method was studied extensively in the context of general 
soliton theory \cite{CHRIST}-\cite{Tomb}, as well as
for Josephson junctions in particular \cite{BJMA}, \cite{TOP_EX}, 
\cite{PER_VOL}, \cite{MC_SCOTT}, \cite{BBJIM}. The basic idea behind it is 
that the field $q$ can be expanded around the soliton solution in a form 
similar to (Eq. (\ref{2_EXP_AR_SOLITON}))
\begin{equation}
\label{2_COL_COO_EXP}
q(x,t)=q_{sol}(x-X(t))+\sum_kc_k\psi_k(x-X(t))\exp(i\omega_kt)\ ,
\end{equation}
where the $c_k$ are the normal modes around the soliton configuration, i.e.
the plasmons, and the sum now does not include the zero mode. The soliton's
center of mass is thus elevated into the role of a dynamic variable at 
the expense of the non-physical zero mode. The Hamiltonian 
(\ref{7_HAM_BETA}) can be now canonically transformed into a form 
involving the new variables. Alternatively, in order to study the dynamics 
of the soliton itself, one can 
retain only the collective coordinate by assuming the form
\begin{equation}
\label{7_RIGID_SOLITON}
q(x,t)=q_{sol}(x-X(t))\ ,
\end{equation}
which means that the soliton is considered to be a rigid object moving with
velocity $\dot X$. This 
assumption is justified when the temperature is much lower than the 
plasmons' energy gap. The plasmons are then treated as a perturbation. In 
this method the particle interpretation of the soliton is clearly seen. 

The collective coordinates can be expressed in an explicit form by using 
the soliton density, $\partial_x q_{sol}$, as a weight function 
\cite{BBJIM}. The center of mass of 
the charge soliton (the position of the excess Cooper pair) is given by
\begin{equation}
\label{7_COL_COO}
X\equiv -{1\over 2\pi}\int x\partial_x q_{sol}\,dx\ ,
\end{equation}
and the conjugate collective momentum by
\begin{equation}
\label{7_COL_MOM}
P\equiv \int\pi_q\partial_x q_{sol}\,dx \ .
\end{equation}
One can check that $X$ and $P$ are indeed canonical variables by 
calculating
their Poisson brackets: $[X,P]_{PB}=1$.
Inserting the soliton configuration (\ref{7_DYN_SOL}) into the definitions
(\ref{7_COL_COO}) and (\ref{7_COL_MOM}) we get the equations of motion of 
a free relativistic particle
\begin{equation}
\label{7_X_EQ}
X=X_0+vt\ ,
\end{equation}
\begin{equation}
\label{2_COL_VEL_EQ}
\dot X=v\ ,
\end{equation}
\begin{equation}
\label{7_REL_P_EQ}
P=\gamma M_0\dot X\ ,
\end{equation}
\begin{equation}
\label{2_COL_MOM_DOT}
\dot P=0\ .
\end{equation}

\subsection{The Dynamic Mass and the Hamiltonian}

The mass that appears in (\ref{7_REL_P_EQ}) is actually the dynamic mass 
of the charge soliton
\begin{equation}
\label{DYN_MASS}
M_d\equiv -{1\over a}\left({2e\over 2\pi}\right)^2L_{kin}
\int\partial_x q_{sol}^2\,dx\ .
\end{equation}
Its value is identical to the rest mass (\ref{7_SOL_MASS}) in the 
limit $L\gg\Lambda_C$, and differs from it by a factor of $2$ in the 
opposite limit $L\ll\Lambda_C$. As we consider here the first limit, we 
denote it by $M_0$ as well.
We can understand the origin of the dynamic mass by observing the way the
charge soliton propagates. Starting from the static distribution of 
voltages on the junctions (Fig. \ref{FIG5}), the center of the charge 
soliton moves 
from its position in the middle of a grain towards one of the neighbouring 
junctions, say the right one, by a charge redistribution in the grains. A 
superposition of charge states in the two adjacent grains is built, and the
(negative) voltage on this junction is reduced. When
the superposition is of states of equal weight, the voltage is zero. As the
motion continues, the charge redistribution increases the weight of the 
charge state on the right grain and the voltage on the junction is 
increased. When the absolute value of this voltage reaches the initial one,
the center of the charge soliton has been shifted by one unit cell, i.e. it
is in the middle of the right grain. One sees that the propagation of the 
charge soliton is determined by the kinetic inductance and not by the 
Josephson one. The dynamic mass leads us, therefore, to the same 
conclusion that we got from the rest mass: the charge soliton is the 
polarization cloud that accompanies the excess Cooper pair that exists in 
the array. 

Transforming now the Hamiltonian (\ref{7_HAM_BETA}) into collective 
coordinates form, we get
\begin{equation}
\label{7_REL_PART_HAM}
H=\sqrt{M_0^2v_C^4+P^2v_C^2}\ ,
\end{equation}
so the energy of the moving soliton is
\begin{equation}
\label{7_FLUXON_PART_ENE}
E=\gamma M_0v_C^2=\gamma E_0\ .
\end{equation}
If we assume the non-relativistic limit, i.e. $v\ll v_c$, the Hamiltonian
describing the soliton as a particle reads
\begin{equation}
\label{7_PART_HAM}
H=M_0v_C^2+{P^2\over2M_0}\ ,
\end{equation}
where now
\begin{equation}
\label{P_EQ}
P=M_0\dot X\ .
\end{equation}
The rest energy term in the Hamiltonian (\ref{7_PART_HAM}) is made out of 
the two charging energies (the last two terms in (\ref{7_HAM_BETA})), 
while the contribution to the kinetic term in (\ref{7_PART_HAM}), comes 
only from the inductive energy (the first term in (\ref{7_HAM_BETA})).
We thus see that the inductive energy, although being the smallest energy 
in the system, is the one that governs the dynamics of the charge soliton.
The independence of the Hamiltonian (\ref{7_PART_HAM}) on $X$ is another
manifestation of the translation invariance of the system. 

\subsection{A Voltage Biased Array}

The collective coordinates can be used to describe a voltage (or a 
time-varying flux) biased array as
well. Introducing the external voltage in the form
\begin{equation}
\label{7_EXT_FLUX}
\dot\Phi_{ext}\equiv -V_{ext}\ ,
\end{equation}
we find that the collective momentum is shifted to
\begin{equation}
\label{7_BIAS_COL_MOM}
P=M_0\dot X+{2\pi\hbar\over L}{\Phi_{ext}\over\Phi_0}\ ,
\end{equation}
and the non-relativistic particle Hamiltonian is
\begin{equation}
\label{7_BIAS_PAR_HAM}
H=M_0v_C^2+{1\over 2M_0}\left(P-{2\pi\hbar\over L}{\Phi_{ext}
\over\Phi_0}\right)^2\ .
\end{equation}
The equations of motion derived from (\ref{7_BIAS_PAR_HAM}) are
\begin{equation}
\label{7_X_EQ_MOT}
\dot X={1\over M_0}\left(P-{2\pi\hbar\over L}{\Phi_{ext}
\over\Phi_0}\right)\ ,
\end{equation}
and
\begin{equation}
\label{7_P_EQ_MOT}
\dot P=0\ .
\end{equation}
Combining the two equations we get
\begin{equation}
\label{7_BIAS_EQ_MOT}
M_0\ddot X=-{2\pi\hbar\over L}{\dot\Phi_{ext}\over\Phi_0}\ ,
\end{equation}
i.e. the external voltage accelerates the charge soliton. The origin of 
this acceleration is simply the electrostatic force exerted on the excess 
Cooper pair by the external voltage. In order that the rigid soliton 
assumption will be valid in this case as well, the external flux must be 
changed adiabatically, or the external voltage should be small enough
\begin{equation}
\label{CL_AD_CON}
\left|{\dot\Phi_{ext}\over\Phi_0}\right|=\left|{V_{ext}\over\Phi_0}\right|
\ll\omega_C\ ,
\end{equation}
which means that $V_{ext}$ should be of the order of $10\,\mu V$ or less. 

\begin{figure}
\centerline{\psfig{figure=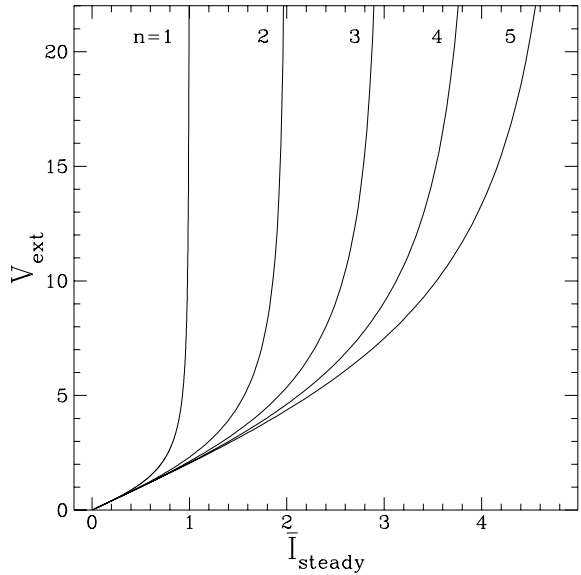,height=8.0cm}}
\caption{I-V characteristic of a voltage-biased dissipative array. Each
branch corresponds to a certain number of charge solitons in the system.
$V_{ext}$ is measured in $\mu V$ and $\bar I_{steady}$ is measured in nA.
The parameters are $R=10\Omega$ and $v_C=10^{-2}c$.}
\label{FIG6}
\end{figure}

When there are Ohmic dissipation processes in 
the array an application of an external voltage results in 
a steady state velocity (or current) of the soliton. The steady state is 
reached when the power gained by the voltage is equal to the power lost via
the dissipation. Using the Hamiltonian (\ref{7_HAM_BETA}), the equation of 
motion (\ref{7_SER_BIAS_RES_EQ_MOT}), and the average current 
(\ref{AVE_CURRENT}), we find that the steady state condition is
\begin{equation}
\label{STEADY_STATE}
V_{ext}=R_{eff}\bar I_{steady}\ ,
\end{equation}
where the effective resistance of the array is constant in the 
non-relativistic case
\begin{equation}
\label{EFF_RES}
R_{eff}\equiv {8\over(2\pi)^2}{L^2\over a\Lambda_C}R
\end{equation}
and is $\bar I_{steady}$-dependent in the relativistic case 
\begin{equation}
\label{EFF_RES_REL}
R_{eff,\,rel}(\bar I_{steady})\equiv 
{8\over(2\pi)^2}{L^2\over a\Lambda_C}\,\gamma(\bar I_{steady})R
\end{equation}
The effective non-relativistic resistance of the 
array is thus increased by about two orders of magnitude, while relativity
increases it further by the $\gamma$ factor. The I-V characteristic of the
array is expected to show saturation branches, where each branch 
corresponds to a certain number of solitons reaching the limit velocity, 
$v_C$ (Fig. \ref{FIG6}).

\section{Quantum Dynamics of the Charge Soliton}

\subsection{The Semi-Classical Expansion}

In this section we study the quantum dynamics of the charge soliton as a 
particle. For this we utilise the semi-classical quantization of the
sine-Gordon theory \cite{COLEMAN}, \cite{Raj}, \cite{CHRIST}, 
\cite{JACKIW}. The expansion parameter is the coupling constant $\beta^2$, 
which was defined in (\ref{7_BETA_SCALE}). In this method the total Fock 
space is taken to be composed of disconnected sectors, each one 
corresponds to different topological boundary conditions, i.e. to a 
different number of solitons in the system. The ground state of each 
sector is the corresponding solitonic configuration. Here we concentrate 
on the one-soliton sector. Due to the translational invariance of the 
system there is, in fact, a degenerate family of eigenstates of the 
position operator, connected by space translations. Higher states are found
by a semi-classical expansion around the ground state. The excitations of 
the first order correspond to the plasmons, and their quantum 
interpretation is as light particles scattering from the static massive 
soliton. These plasmons are, thus, the fundamental quanta of the theory. 
The degeneracy of the states is completely removed in the second order, as 
the position eigenstates are replaced by momentum eigenstates, and the 
translation invariance of the theory is recovered on the quantum level. 
The semi-classical expansion breaks down when $\beta^2\ge 2$, where the 
soliton becomes lighter than the plasmons. The soliton then takes the role 
of the fundamental quantum, and loses its correspondence to the classical 
particle configuration (in a sense it becomes `too' quantum). Since the 
typical value of $\beta^2$ is $10^{-1}$, we can use the expansion for the 
array. The parameter $\beta^2$ can be expressed in the form \cite{PHD}
\begin{equation}
\label{BETA_SEMI}
\beta^2=\sqrt{E_L\over E_{C_0}}={R_Q\over Z_{LC}}\ .
\end{equation}
Comparing Eq. (\ref{BETA_SEMI}) with (\ref{IMP_CON}) and (\ref{E_L_E_C_0}),
we see that the condition for using the semi-classical expansion, 
$\beta^2\ll 1$ is identical to the impedance condition. 
This is not a surprise, as the impedance
condition is the one that enables us to decouple the junctions 
quantum mechanically. Our model of the charge soliton as a classical 
configuration is thus self-consistent.

However there are several differences between the system we study and the 
field theoretical model. First of all, the array is very long (compare
to $\Lambda_C$), but finite. Apart from a slight distortion to the 
soliton's shape
that we neglect, the finiteness means that solitons can enter and leave the
array, and also get reflected from the edges. To avoid this situation, we
consider a ring-shaped array.
Second, since the gap in the plasmons' spectrum is of the
order of one Kelvin, their population can be made negligible if the
temperature is kept below the gap. Thus we can discard all the plasmons' 
contribution to the dynamics. This assumption 
is equivalent to the rigid soliton assumption (\ref{7_RIGID_SOLITON}).
A finite
population of plasmons can be considered as an internal environment which
produces a phase breaking mechanism \cite{DEPH}. We comment on this 
dephasing process at the end of this section.
Another different feature, is that we couple the array to an external
(classical) flux source as a gauge coupling, and study the quantum
dynamics of the soliton in response to this source. Finally, the array we 
study deviates from the ideal sine-Gordon model by its discreteness, by the
exact form of the potential energy,
by structural inhomogeneities and disorder and by quasi-particle
tunneling. The effects of these deviations from the ideal model are 
discussed below. 

\subsection{Persistent Motion of the Charge Soliton}

In the presence of an external flux, $\Phi_{ext}$, the assumption of 
rigidity leads to the following non-relativistic quantum Hamiltonian for a 
ring-shaped array of serially coupled Josephson junctions
\begin{equation}
\label{7_Q_BIAS_PAR_HAM}
\hat H=M_0v_C^2+{1\over 2M_0}\left(\hat P-{2\pi\hbar\over L}{\Phi_{ext}
\over\Phi_0}\right)^2\ .
\end{equation}
Higher orders contributions to the energy give rise to quantum corrections
to the soliton's rest mass \cite{DHN_74b}. The renormalized
mass in the array language (up to the order of $\beta^0$) is
\begin{equation}
\label{3_REN_MASS}
M_{0\,ren}={8\over\Lambda_C}{\hbar\over 2\pi\beta^2v_C}
\left(1-{\beta^2\over 4}\right)=M_0\left(1-{\beta^2\over 4}\right)\ .
\end{equation}
However since $\beta^2$ is small we can use $M_0$ instead of $M_{0\,ren}$. 
As we have discussed in the previous section, the Hamiltonian is $\hat X$ 
independent due to the homogeneity of the array. Thus it commutes with the
collective momentum operator,
$\hat P$, and the eigenstates are collective momentum eigenstates with a 
discrete set of eigenvalues, $p_N=\hbar k_N$ determined by the periodic 
boundary conditions
\begin{equation}
\label{P_QUAN}
k_N={2\pi\hbar\over L}N\,;\ \ \ \  N=0,\pm 1,\pm 2,...\ .
\end{equation}
The energy spectrum is discrete, too, and is given by (neglecting
the constant term $M_0v_C^2$)
\begin{equation}
\label{ENE_LEV}
E_N={1\over 2M_0}\left({2\pi\hbar\over\Phi_0L}\right)^2
(\Phi_0N-\Phi_{ext})^2\ .
\end{equation}
Defining an effective inductance by
\begin{equation}
\label{EFF_IND}
L_{eff}\equiv M_0\left({\Phi_0L\over2\pi\hbar}\right)^2
\end{equation}
($L_{eff}\approx 10^{-5}\,$ Henry), the energy levels can be 
expressed in the form of inductive levels
\begin{equation}
E_N={1\over2L_{eff}}(\Phi_0N-\Phi_{ext})^2\ .
\end{equation}
The inductive form of the energy levels suggests the interpretation of
$N$ as the number of flux quanta that has tunneled outside or inside 
the ring through one of the junctions. The quantization of $\hat P$ is, 
therefore, the statement that only an integral number of flux quanta 
can tunnel in or out of the ring. However, the conservation of the 
momentum means that there can be no flux tunneling in an homogeneous 
array, i.e. the external flux is completely screened. 

\begin{figure}
\centerline{\psfig{figure=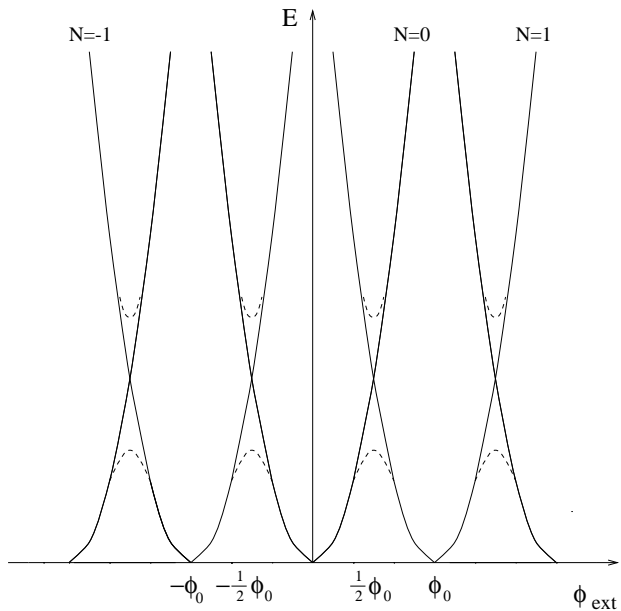,height=8.0cm}}
\caption{The spectrum of a quantum charge soliton in a 1-D
ring-shaped array of serially coupled Josephson junctions as a function
of an external flux, $\Phi_{ext}$. In an ideal ring the spectrum consists
of inductive energy parabolas without a possibility of crossing at
the intersection points. When there is some inhomogeneity in the ring (e.g.
due to disorder), gaps are open at the intersection points, and the
spectrum develops into energy bands.}
\label{FIG7}
\end{figure}

The spectrum of the charge soliton's Hamiltonian (\ref{ENE_LEV}) is 
periodic with respect to the external flux with a period $\Phi_0$. 
It is composed of a set of parabolas centered at $\Phi_{ext}=N\Phi_0$. 
Each parabola intersects its two adjacent parabolas at $(N+1/2)\Phi_0$
(Fig. \ref{FIG7}). The current along the array is given by
\begin{equation}
\label{PERS_CUR}
\langle I\rangle=-{\partial E_N\over\partial\Phi_{ext}}
={1\over L_{eff}}(\Phi_0N-\Phi_{ext})\ .
\end{equation}
It is proportional to the expectation value of the velocity of the
charge soliton
\begin{equation}
\label{EXP_VEL}
\langle\dot X\rangle={L\over2\pi\hbar}{\partial E_N\over\partial N}
={L\over 2e}\langle I\rangle\ .
\end{equation}
This is the quantum version of the relation (\ref{AVE_CURRENT}). 
We see that the external flux induces a persistent motion of the 
charge soliton, which is manifested 
in a persistent current along the array. As was shown above, no net 
number of flux quanta can tunnel in or out of the junction. However, 
during the motion of the soliton one can think of flux quanta flowing 
in and out of the array through the junctions, thus forming a flux loop 
around the moving center of the soliton. (A similar idea for 2-D 
superconducting films was given in \cite{Film_Duality}.) 
This interpretation is dual to the interpretation of the fluxon in a 
long Josephson junction as a (charge) current loop. 
The charge soliton's persistent current has the same origin 
as the persistent current of electron in a metal ring \cite{PER_CUR}. 
It is a manifestation of the Aharonov-Bohm effect \cite{AB} of a 
charged particle encircling a flux tube, and its persistency is due 
to the particle being in an exact eigenstate of the system. 
However, in contrast to the electron, the charge soliton is a macroscopic
particle ($\Lambda_C\approx 100\,\mu m$), so the possibility that it 
exhibits quantum effects is very intriguing.
The quantum behaviour of the charge soliton is dual to the quantum 
behaviour of the fluxon in a long Josephson junction \cite{PER_VOL}.
The latter exhibits a persistent motion in 
response to an external bias charge, which is the manifestation of the 
Aharonov-Casher effect 
\cite{AC}. Being a magnetic particle, this motion results in a 
persistent voltage across the junction.

A weak spatial inhomogeneity in the array, e.g. non-identical grains
or junctions or disordered grains, gives an additional 
$\hat X$ - dependent term in the Hamiltonian (\ref{7_Q_BIAS_PAR_HAM}). 
The momentum is not conserved anymore, and flux quanta can tunnel across 
the array, reflecting in the spectrum by gaps which are opened at the 
intersection points of the parabolas (see Fig. \ref{FIG7}). If the array 
is now adiabatically biased by a time 
varied flux source, the persistent current oscillates as a function 
of $\Phi_{ext}$ with a period $\Phi_0$. In each period a flux quantum
tunnels across the array. This tunneling creates a current in the
inverse direction to the existing current, thus eliminating the net
current and reducing the energy. Since the energy bands are exact 
eigenstates, the tunneling process is a coherent one. When the 
external flux is not equal to an integral number of flux quanta, the 
quantum state of the array is a superposition of two flux quantum 
states. The amplitude of a persistent current of one charge soliton
decreases as the amount of inhomogeneity increases. The maximal 
amplitude, corresponding to a vanishing amount of inhomogeneity, is 
of the order of $0.1\,$nA. 

\subsection{Other Quantum Effects}

The quantum nature of the charge soliton can be revealed in transport
phenomena as well. For instance, if solitons are sent through a 
ring-shaped array connected to two leads (all consist of serially coupled 
Josephson junction), and the dephasing mechanisms are suppressed, we expect
that they will split into partial waves propagating along the two arms of 
the ring. The partial waves will then interfere at the outgoing leads, with
the interference pattern being dependent on the length of the arms and on 
an external flux applied through the center of the ring. 
The transmission of quantum charge solitons through the ring is thus 
expected to show oscillations as a function of the external flux and 
of the optical path similar to the $h/e$ oscillations in the transmission 
of electrons through a metal ring \cite{h/e}, and in analogy to the 
transmission of fluxons through a Josephson junction ring 
\cite{INTERFERENCE}.

\subsection{Dephasing Mechanisms}
The quantum phenomena described above were a
consequence of the fact that in our approximation the Hamiltonian
(\ref{7_Q_BIAS_PAR_HAM}) was a one particle Hamiltonian. Thus, even in
the presence of a weak inhomogeneity, the degree of freedom 
associated with the charge soliton's center of mass ($X_0$) can
maintain its quantum coherence. In order to make the model more 
realistic, one should take into account interactions between the
soliton and other degrees of freedom. These interactions can produce,
in principle, phase breaking mechanisms. Whenever the phase breaking
length, defined as the length over which the soliton's phase has an 
uncertainty of $2\pi$, is smaller than the length of the array, the 
quantum phenomena exhibited by the charge soliton will be 
suppressed. As in the case of the fluxon in a long Josephson 
junction \cite{DEPH}, we can distinguish between internal and external 
dephasing mechanisms. The internal mechanism
is due to the interaction between 
the charge soliton and the other degrees of freedom of the junction, 
i.e. the plasmons. When the sine-Gordon model is exact and continuous,
the system is completely integrable and the soliton is decoupled from the 
plasmons. Nevertheless, it has been shown in the context of 
the fluxon in a long Josephson junction \cite{DEPH} that there is a 
possibility of dephasing in this case as well. In order to avoid this
dephasing, the temperature should be below the plasmons' energy gap. 
In the context of the charge soliton, where the sine-Gordon model
is only an approximation and the system is discrete, we expect that the
plasmons give rise to a stronger dephasing due to their inelastic 
interaction with the soliton. From the study of the discrete sine-Gordon 
model it is known that the rest energy of a soliton whose center resides 
in a junction is higher than the rest energy for a soliton whose center 
resides in the middle of a grain \cite{Pey_Krus_Dis_sG}. Thus the soliton 
propagates in a periodic potential and not in a flat one. This 
deviation from the continuum model produces a coupling between the 
plasmons and the soliton. The soliton can emit or absorb plasmons 
\cite{Currie_Dis_SG}, 
\cite{Pey_Krus_Dis_sG}, and the circulating soliton can become phase 
locked with this plasmons \cite{Dis_sG_Ust_Cir_Bor}. This effect has been 
recently observed for the fluxon in the discrete long Josephson junction 
\cite{Her_Dis_Lock}. We expect that similar phenomena occur in the system
we study here when the continuum condition (\ref{7_CON_LIM_LAM}) does 
not hold. Apart from producing a phase breaking length, these phenomena
will affect the classical dynamics as well, for instance by creating 
resonances in the I-V characteristic. The influence of both the
discreteness of the array and the deviation from the exact sine-Gordon
model on the classical and quantum mechanical dynamics of the charge 
soliton should be studied further.

The most important external dephasing mechanisms are due to interaction 
with quasi-particles, which was neglected in our model. 
Since the bulk superconductors
energy gap, $\Delta$, is typically of the same order or higher than 
the plasmons' energy gap, the condition needed to suppress the thermal 
activation of the plasmons is sufficient to suppress the thermal activation
of the quasi-particles. The effects of thermal quasi-particles will
be studied elsewhere.
The quasi-particles can destroy the quantum coherence of the array
in another way, which is temperature independent. The complete spectrum
of a single junction includes charging energy parabolas
associated with quasi-particles as well, which are separated in the
charge axis by $e$. Excitation of quasi-particles leads to transitions
between these parabolas, thus destroying the quantum coherence of $q$.
This effect can be neglected if the charging energy of the quasi-particles
plus the superconducting energy gap is larger than the Cooper pairs
charging energy, i.e.: $2\Delta+e^2/8C>e^2/2C$ or ${32\over 3}\Delta>E_C$. 
Since $E_C$ should be smaller than $\Delta$ for the existence of the 
Josephson effect, this condition is met automatically.

\section{Summary}

We have studied a 1-D array of serially coupled Josephson junctions in the
limit when the kinetic inductance of the superconducting grains dominates 
over the Josephson inductance. In this case the array is described by 
variables which are defined on the junctions and not on the grains. We have
shown that the large kinetic inductance decouples the junctions quantum 
mechanically. As a result each junction is characterised by a periodic 
charging energy. This periodic energy, when combined with the inductive 
energy of the grains and the charging energy between the grains and the 
substrate, gives rise to a model with topological solitons excitations. 
Thus we have found that an excess Cooper pair in the array creates a charge
soliton via polarization of the superconducting grains. The charge soliton 
is a dual topological excitation to the fluxon in a long Josephson 
junction. We have studied the classical dynamics of the charge soliton, and
shown that in the presence of dissipation and an external voltage the I-V 
characteristic of the array should consist of saturation branches 
corresponding to the number of charge solitons in the array. We have 
quantized the charge soliton semi-classically, showing that this 
quantization is consistent with the large kinetic inductance. We have found
that a quantum soliton in a flux-biased ring-shaped array is expected to 
show persistent motion, manifested in a persistent current. A weak 
inhomogeneity in the array gives rise to a coherent current oscillations. 
These phenomena, which are usually associated with electrons (or Cooper 
pairs) suggests that the quantum charge soliton can be considered as a 
macroscopical quantum object. Finally, we have discussed possible internal 
and external dephasing mechanisms of the charge soliton. These mechanisms 
deserve a future study.

{\bf Acknowledgment}:
We like to thank G.~D.~Guttman, A.~Shnirman, D.~B.~Haviland, P.~Delsing, 
R.~Fazio, F.~Guinea, A.~V.~Ustinov, B.~A.~Malomed and S.~E.~Korshunov 
for very fruitful discussions. One of us (Z.~H.) is supported by the 
MINERVA fellowship. This research was supported
in part by the Wolfson foundation via the Israeli Academy of Sciences,
and by the DFG within the research program of the Sonderforschungbereich 
195.

\end{document}